\documentclass[12pt]{iopart}
\usepackage{iopams}
\usepackage{graphicx}
\usepackage{amssymb,amsthm}

\newcommand{\ba}{\boldsymbol{\alpha}}
\newcommand{\bx}{\boldsymbol{\xi}}
\newcommand{\bex}{\boldsymbol{x}}
\newcommand{\by}{\boldsymbol{y}}
\newcommand{\bet}{\boldsymbol{\eta}}
\begin{document}

\title{Gravitational lensing by eigenvalue distributions of random matrix models
 }

\author{Luis Mart\'{\i}nez Alonso}

\address{Departamento de F\'{\i}sica Te\'orica II,
                               Facultad de Ciencias F\'{\i}sicas,
                               Universidad Complutense,
                               28040 Madrid, Spain}
\ead{luism@fis.ucm.es}

\author{Elena Medina}

\address{Departamento de Matem\'aticas,
                      Facultad de Ciencias,
                      Universidad de C\'adiz,
                      11510 Puerto Real, Spain}
\ead{elena.medina@uca.es}

\vspace{10pt}
\begin{indented}
\item[September 2017]
\end{indented}

\begin{abstract}
We propose to use eigenvalue densities of unitary random matrix ensembles as mass distributions in  gravitational lensing. The corresponding lens equations  reduce to algebraic equations in the complex plane which  can be treated analytically.  We prove that these models can be applied to describe lensing by systems of edge-on galaxies. We illustrate our analysis with the Gaussian and the quartic unitary matrix ensembles.
\end{abstract}

\pacs{02.10.Yn, 02.90.+p}
%
%
\submitto{\CQG}
%
%
%

\section{Introduction}

The trajectory of  light from distant sources is perturbed by foreground  distributions of matter such as galaxies, groups and clusters. This phenomenon is called gravitational lensing and it has  important  applications in cosmology  \cite{SH92,NA96}.
The deflection angle  $\widehat{\alpha}$ of a light ray passing at a minimum distance $\xi$ (impact parameter) of a point-like mass $M$  is  given by the Einstein's formula 
\begin{equation}\label{ei}
\widehat{\alpha}=\frac{4 GM}{\xi},
\end{equation}
see e.g. Chapter 4 of  Ref. \cite{SH92} and Figure 1, where $G$ is the constant of gravity. In general,  the deflection angle  is defined as the vector  $\widehat{\ba}={\bf e}_i-{\bf e}_f$ where ${\bf e}_i$ and ${\bf e}_f$ are the initial and final  unit tangent vectors to the light ray trajectory from the source to the observer, respectively.  For  thin lenses with continuous mass distributions and  weak  gravitational fields, the deflection angle is the sum of the deflections due to  each mass element of the lens. Thus $\widehat{\ba}(\bx)$ is a function $\widehat{\ba}(\bx)$ of the  impact vector $\bx$  in the lens plane \cite{SH92},
\begin{equation}\label{alp}
\widehat{\ba}(\bx)=4G\int_{\mathbb{R}^2} \frac{\bx-\bx'}{|\bx-\bx'|^2}\Sigma(\bx') d^2 \bx',
\end{equation}
where $\Sigma(\bx) $ is the surface mass density obtained by projecting  the volume mass distribution of the deflector onto the lens plane. 
Moreover, there is a geometrical relation (see Chapter 4 of  Ref. \cite{SH92})  between the source position $\bet$ and the impact vector $\bx$ of the ray  in the lens plane 
\begin{equation}\label{lene}
\bet=\frac{D_s}{D_d}\bx-D_{ds}\widehat{\ba}(\bx),
\end{equation}
where $D_s$, $D_d$ and $D_{ds}$ are the distances from the observer to the source plane, from the observer to the lens, and from the lens to the source plane, respectively (see the diagram of Figure 1).  In effect, we have a correspondence between vectors $\bx$ in the lens plane and vectors  $\bet$ in the  source plane. The problem of gravitational lensing is the inversion of this correspondence, i.e. to determine the image positions $\bx$ for a given source position $\bet$. Images $\bx$ outside (inside) the mass distribution $\Sigma(\bx) $  are referred to as {\em bright} ({\em dim}) images. 

\begin{center}
\begin{figure}[ht]
\centering
\includegraphics[width=10cm]{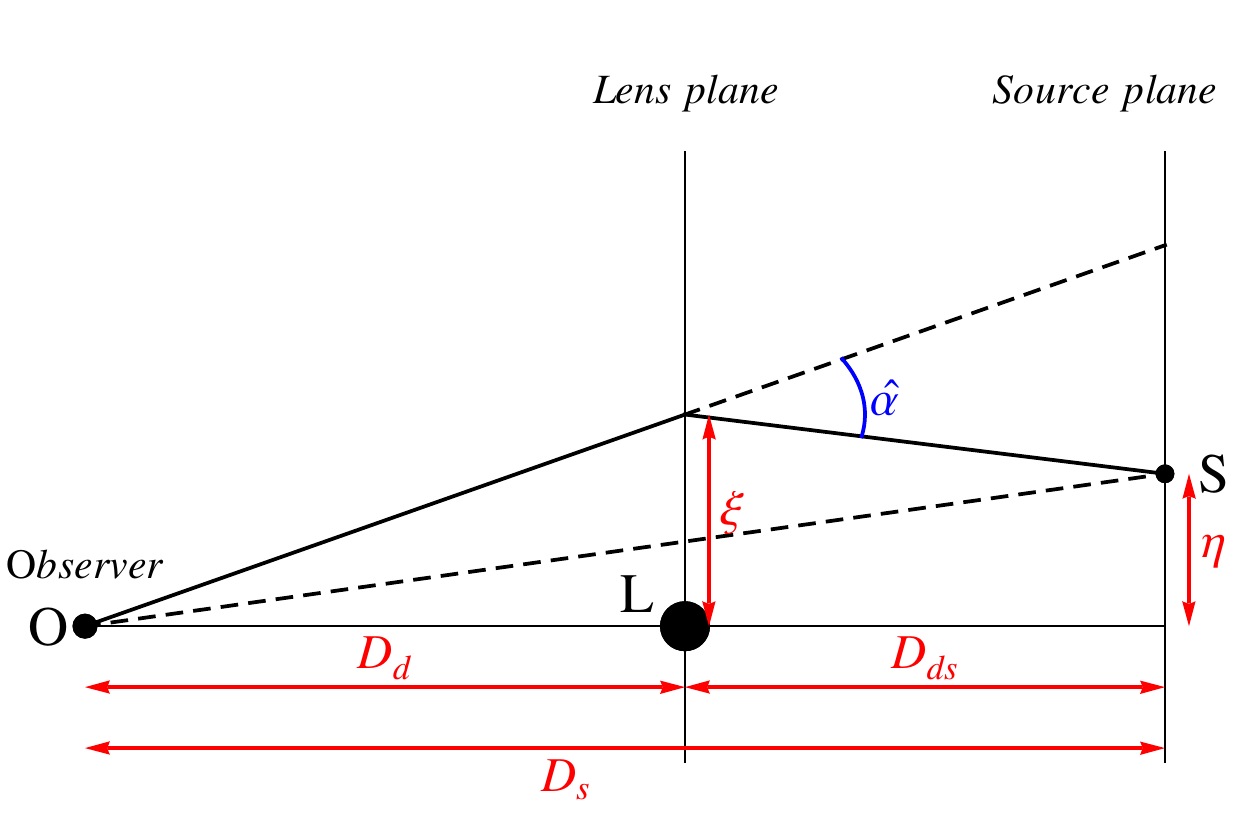}
\caption{A lensing system with a source  $S$, a lens $L$ and an observer $O$. The lens  deflects a light ray emanating from the  source  by an angle $\widehat{\alpha}$.}
\end{figure}
\end{center}

  The computation  of the image positions  for a given source is a difficult   problem \cite{BL14}, which in general can be solved only numerically. Even the two point-mass lens  \cite{SC86}  is already very complicated to study analytically.  Hence,  it is interesting to find lensing systems allowing partial or total  analytic treatment. Among the  possible candidates are the  mass distributions  defined by measures with \emph{minimal support} (consisting of  a finite set of isolated points or/and a finite number of compact analytic curves). Measures of this type arise in the study  of   families  of measures producing the same external gravitational potential and are known as \emph{ mother bodies} in geophysics \cite{ZI68}. They have been mathematically defined in several forms, see e.g. \cite{GU98,SA05,BO16}. The concrete notion of mother body measure  used in this paper is given in the next section.

The present work is  devoted to  lensing models in which the  mass distributions are determined by  the eigenvalue densities  of unitary random matrix ensembles   in the limit of large matrix dimension \cite{ME91,DI95,BL08}. These mass distributions  are  non-trivial examples of  mother bodies supported on a finite number of real intervals (cuts). The corresponding lens equations  turn out to reduce to algebraic equations and exhibit interesting classes of explicit analytic solutions. In particular, we propose to apply  these models to describe  gravitational  lensing by  disk galaxies seen edge-on (see \cite{KU06} for a catalog of this type of galaxies). Thus an $n$-cut eigenvalue density describes a lensing system of $n$ coplanar edge-on galaxies. Consequently,  phase transitions of eigenvalue distributions  in which the number of cuts  changes   \cite{BL08, AL10,MA15} represent splitting-merging processes  of edge-on galaxies. Moreover, due to their mother body character, the eigenvalue distributions can be also applied to determine the bright images of more general mass distributions.  We illustrate our analysis with two well-known random matrix models: the Gaussian and the quartic models. We study their  eigenvalue distributions as mother bodies supported on elliptic domains and formulate  their associated lensing models by edge-on galaxies. We consider the corresponding  lens equations and derive  explicit expressions for wide classes of  dim and bright images. We also  include some exact calculations of time delays.

  The layout of this paper is as follows. Section~\ref{sec:ei} concentrates on the main properties of the eigenvalue distributions of unitary ensembles of random matrix models and formulates the corresponding lens models. Then we  discuss  these distributions as mother body measures  and  edge-on disk galaxies. Sections~\ref{sec:gaus} and~\ref{sec:quar}  contain our  main explicit results. Section~\ref{sec:gaus} is devoted to the lens model of  the Gaussian eigenvalue distribution and Section~\ref{sec:quar} deals with the lens model corresponding to the unitary ensemble of random matrices with a quartic potential. After some concluding remarks on several open questions, the paper ends with one appendix  where it is proved that the Gaussian and the quartic models are mother bodies on elliptic domains.

\section{Gravitational lensing by eigenvalue distributions\label{sec:ei}}

\subsection{Complex formulation of the lens equation \label{sec:leq}}
 It is convenient to write the lens equation (\ref{lene}) in dimensionless form. Thus, we  introduce dimensionless vectors $\bex=\bx/\xi_0$ and $\by=\bet/\eta_0$, where $\xi_0$ is a fixed length scale (whose choice depends on the particular problem studied)  and $\eta_0=\xi_0 D_s/D_d$. Then  we have
\begin{equation}\label{alpd}
\by=\bex-\int_{\mathbb{R}^2} \frac{\bex-\bex'}{|\bex-\bex'|^2}\kappa(\bex') d^2 \bex',
\end{equation}
where $\kappa(\bex)=4 G D_d D_{ds}\Sigma(\xi_0 \bex)/D_s$.
If we now represent the two-dimensional vectors $\bex$ and $\by$  by complex numbers $z$, $w$, and complex conjugate  the resulting equation, we obtain \cite{ST97} 
\begin{equation}\label{len}
\overline{w}=\overline{z}-\omega(z),
\end{equation}
where  $\omega(z)$ is  the Cauchy transform  (in the principal value sense)
\begin{equation}\label{cau}
\omega(z)=\int_{\mathbb{C}} \frac{d \mu(\zeta)}{z-\zeta},
\end{equation}
of the  mass distribution $d\mu=\kappa(\zeta) d^2 \zeta$.  In order to avoid the singularity at $\zeta=z$ the integral (\ref{cau}) is assumed to be defined in the principal value sense, (i.e. we integrate on the domain $ |\zeta-z|>\epsilon$ of the $\zeta$ plane  and then we take the limit as $\epsilon\rightarrow 0$). Equation (\ref{len}) provides a quite useful  complex formulation of the lens equation which allows us to apply various tools of complex analysis.

\subsection{Eigenvalue distributions of matrix models \label{sec:crit}}

We consider  random matrix models of $N \times N$ Hermitian matrices $M$  with Boltzmann weights of the form $\exp (-2 N V(M)) $ where  \emph{the potential function} $V(x)$ is a real polynomial function  of even degree $V(x)=\sum_{i=1}^{2p}t_i x^i$  with $t_{2p}>0$ (unitary ensembles)  \cite{ME91,DI95,BL08}. The associated unit normalized eigenvalue density  in the limit $N\rightarrow \infty$ is supported on a finite  union of disjoint real intervals
\begin{equation}\label{pol}
\Gamma=\cup_{i=1}^n [a_i,b_i],
\end{equation}
with
\begin{equation}\label{norm1}
\int_{\Gamma} \rho(x)\, dx=1,
\end{equation}
and is of the form
\begin{equation}\label{un}
\rho(x)=p(x) \sqrt{\prod_{i=1}^n(x-a_i)(b_i-x)},
\end{equation}
for a given polynomial  $p(x)$. 

 The density $\rho(x)$ determines a  mass distribution $d \mu_e=\rho(x) d x$ with the one-dimensional support $\Gamma$. From the point of view of the applications to gravitational lensing the main property of $\mu_e(z)$  is that its  Cauchy transform
  \begin{equation}\label{caud}
\omega_e(z)=\int_{\Gamma} \frac{\rho(x) d x}{z-x},
\end{equation}
 satisfies  a quadratic equation \cite{DI95,BL08}
\begin{equation}\label{alg}
\Big(\omega_e(z)-V'(z)\Big)^2=P(z),
\end{equation}
where  $P(z)$  is a polynomial such that ${\rm deg} \,P=2 \,{\rm deg} \,V-2$. As a consequence the function  $\omega_e(z)$  can be explicitly calculated in terms of $V$ and $P$. Indeed, from (\ref{alg})  and taking into account that $\omega_e(z)\sim 1/z$ as $ z\rightarrow \infty$ we have that outside $\Gamma$
\begin{equation}\label{alg2}
\omega_e(z)= V'(z)-\sqrt{P(z)} , \quad z\in
\mathbb{C}\setminus \Gamma,
\end{equation}
where  the branch of $\sqrt{P(z)}$ is determined by the condition
\begin{equation}\label{algc}
V'(z)-\sqrt{P(z)}=\frac{1}{z}+\mathcal{O}(1/z^2),\quad z\rightarrow \infty.
\end{equation}
The expression of $\omega_e(z)$ on $\Gamma$ can be derived from the  boundary values   of $\omega_e(z)$ on both sides of $\Gamma$
\begin{equation}\label{bv}
\omega_{e\pm}(x)=\lim_{\epsilon\rightarrow +0}\omega_{e}(x+\pm i \epsilon), \quad x\in \Gamma.
\end{equation}
 Thus, since the integral (\ref{caud}) is defined in the principal value sense,   the Sokhotsky-Plemelj formulas \cite{he93,AB03} 
\begin{equation}\label{sp}
\omega_{e\pm}(x)=\omega_{e}(x)\mp i \pi \rho(x),\quad x\in \Gamma,
\end{equation}
imply that
\begin{equation}\label{alg2c}
\omega_e(x)= \frac{1}{2}\Big(\omega_{e+}(x)+\omega_{e-}(x) \Big), \quad x\in \Gamma.
\end{equation}
Hence, as $\sqrt{P(z)}$ has opposite boundary values on both sides of  $\Gamma$, from (\ref{alg2}) and (\ref{alg2c}) we get
\begin{equation}\label{alg3}
\omega_e(x)= V'(x) , \quad x \in \Gamma.
\end{equation}
Therefore (\ref{alg2}) and (\ref{alg3}) provide an explicit  characterization of the Cauchy transform $\omega_e(z)$ on the whole complex plane.

Similarly, it follows from (\ref{alg2}) and (\ref{sp}) that  the eigenvalue  density $\rho(x)$ is given by
\begin{equation}\label{den}
\rho(x)=\frac{1}{\pi} \Big| \sqrt{P(x)} \Big |,\quad x\in \Gamma.
\end{equation}
Furthermore, it can be proved \cite{DI95,BL08}  that $\rho(x)$ represents an equilibrium configuration  in the sense  that the  total potential function
\begin{equation}\label{cou}
{\rm Re}V(z)+U(z), \quad z\in \mathbb{C},
\end{equation}
where $U(z)$ denotes the logarithmic potential 
 \begin{equation}\label{log}
U(z)=-\int_{\Gamma} \ln |z-x|\, \rho(x) d x,\quad z\in \mathbb{C},
\end{equation}
 is constant on  $\Gamma$
\begin{equation}\label{cou1}
V(x)+U(x)=U_0, \quad x\in \Gamma.
\end{equation}

\subsection{Lensing by
eigenvalue  distributions}

Henceforth we consider mass distributions  of the form
\begin{equation}\label{qc}
 \mu=m\,\mu_e,
\end{equation}
where  the parameter $m>0$ represents  the total mass of $\mu$. From (\ref{alg2}) and (\ref{alg3}) we have that
these  distributions have an explicit Cauchy transform  $\omega=m\omega_e$
and  the lens equation  is determined by the functions $V(z)$ and $P(z)$ as follows
\begin{equation}\label{ll}
\left\{\begin{array}{ll} w=x-mV'(x) ,\quad  \mbox{if $x\in \Gamma$ }, \\\\  \overline{w}=\overline{z}-mV'(z)+m\sqrt{P(z)} , \quad \mbox{if $z\in
\mathbb{C}\setminus \Gamma$.} \end{array}\right.
\end{equation}

An important parameter in gravitational lensing is the time it takes the light to travel  from the source to the observer \cite{SH92}. In Eq.(5.45) of Ref. \cite{SH92} it is proved that up to an additive constant  the excess travel time for a light ray with source $w$ crossing
 the lens plane at $z$, relative to an undeflected ray  is
 proportional to 
\begin{equation}\label{arr}
 \tau(z)=\frac{1}{2}|z-w|^2+m\,U(z),
\end{equation}
where $U(z)$ is the logarithmic potential determined by the mass distribution $\mu_e$. We will henceforth refer to $\tau(z)$ as the \emph{time delay}.
In particular, from (\ref{cou1}) and (\ref{arr}) we have that given a dim image $z=x\in \Gamma$
\begin{equation}\label{arr1}
\tau(x)=\frac{1}{2}|x-w|^2+ m\Big( U_0-V(x)\Big),\quad x \in \Gamma.
\end{equation}
Hence  (\ref{arr1})  implies 
 that  the relative  time delay between  two dim images  for a given source $w$ is proportional to
\begin{equation}\label{atrr1}
\tau(x_2)-\tau(x_1)=\frac{1}{2}\Big(|x_2-w|^2-|x_1-w|^2\Big)+ m\Big( V(x_1)-V(x_2)\Big).
\end{equation}

  \subsection{Mother bodies }

Let  $D$ be a bounded domain of the complex plane and $\mu_D$ a measure  with support $\overline{D}$. We may think of the pair $(D,\mu_D)$ as  a  planar body with mass distribution $\mu_D$. Then  another measure $\mu$ is said to be a mother body for $(D,\mu_D)$ if \cite{GU98,SJ06}:
\begin{enumerate}
\item  The support of $\mu$ is  a finite set of curve segments or/and points contained in
$\overline{D}$ such that each connected component of $\mathbb{C}\setminus {\rm supp}\mu$ does not disconnect any part of $D$ from the complement of $D$.
\item The  gravitational potentials of   $\mu$ and  $\mu_D$ coincide on $\mathbb{C}\setminus \overline{D}$.
\end{enumerate}

Since the Cauchy transform and the logarithmic  potential of a measure satisfy $\omega(z)=-2\,\partial_z U(z)$ for $z$ outside the support of the measure,
we have the following important property: A mother body  measure for  $(D,\mu_D)$  produces  the same  bright images as  $\mu_D$ in $\mathbb{C}\setminus \overline{D}$.

The measures  $\mu$ determined by the eigenvalue distributions  (\ref{pol})-(\ref{un})  satisfy the condition  (i) of the above definition for any domain $D$  such that $\Gamma\subset D$. Hence these measures are mother bodies of all the planar bodies $(D,\mu_D)$ with the same gravitational potential as $\mu$ on $\mathbb{C}\setminus \overline{D}$.

Another  definition of mother body  was recently formulated in  \cite{BO16}. According to this alternative definition all measures with Cauchy transform which coincide a.e. in $\mathbb{C}$ with an algebraic function are mother bodies. Then, as a consequence of  the loop equation (\ref{alg}), the measures  determined by the eigenvalue distributions  of unitary ensembles of matrix models are mother bodies in the sense of \cite{BO16}.

\subsection{Edge-on  galaxies}

 The lensing model corresponding to a $n$-cut eigenvalue distribution (\ref{qc})  can be applied to describe  a  system  composed of  $n$  edge-on   disk galaxies \cite{KU06}. These galaxies must be  located on a common plane ($XY$-plane) orthogonal to the lens plane  and such that their  supports  project to  the intervals $[a_i,b_i]$ in the lens plane with  density $\rho(x)$ (Fig. 2).  For example this is the case if the  galaxies have  uniform mass distributions  supported  on the  domains $\mathcal{D}_i$ in the plane $XY$ bounded by the set of closed curves
\begin{equation}\label{proa}
Y^2=\frac{S_i^2}{4} \Big(\frac{m}{ m_i}\Big)^2\,\rho(X)^2 ,\quad a_i\leq X\leq b_i,\quad i=1,\ldots,n,
\end{equation}
where $S_i$ and $m_i$  are the area and the mass of $\mathcal{D}_i$, respectively.

\begin{center}
\begin{figure}[ht]
\centering
\includegraphics[width=6cm]{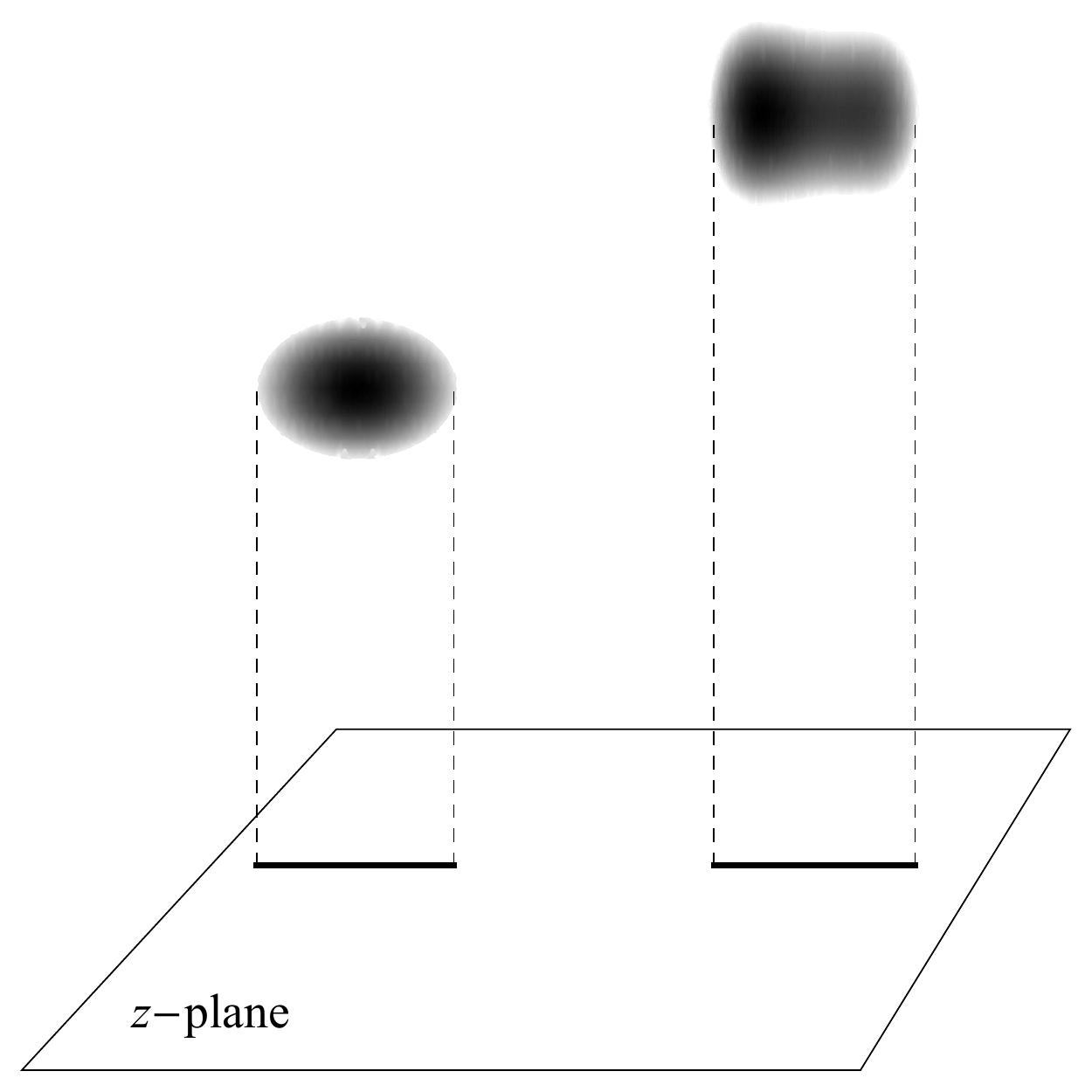}
\caption{The projected supports of coplanar edge-on galaxies is a union of intervals  along a straight line  on the lens plane.}
\end{figure}
\end{center}

\section{The Gaussian model \label{sec:gaus}}

The Gaussian matrix model is determined by the potential function
\begin{equation}\label{gau}
V(z)=\frac{z^2}{a^2},
\end{equation}
where $a>0$ and constitutes a basic model in random matrix theory \cite{ME91,DI95,BL08}.
 In this case there are only one-cut eigenvalue distributions, the  function $P(z)$ in (\ref{alg}) is
\begin{equation}\label{ere}
P(z)=\frac{4}{a^4}(z^2-a^2),
\end{equation}
and the unit normalized mass density takes the form  of the Wigner's semi-circle law (Fig. 3)
\begin{equation}\label{gaud}
\rho(x)=\frac{2 }{\pi a^2}\sqrt{a^2-x^2},\quad -a\leq x\leq a.
\end{equation}
Thus, the  lens equation for a  Gaussian lensing model of mass $m$ is
\begin{equation}\label{gaue1}
\left\{\begin{array}{ll}  w=(1-p) x ,\quad  -a\leq x\leq a, \\\\  \overline{w}=\overline{z}-p(z-\sqrt{z^2-a^2}) , \quad z\in \mathbb{C}\setminus [-a,a],\end{array}\right.
\end{equation}
where
\begin{equation}\label{p1}
p=\frac{2m}{a^2}.
\end{equation}
\begin{center}
\begin{figure}[ht]
\centering
\includegraphics[width=6cm]{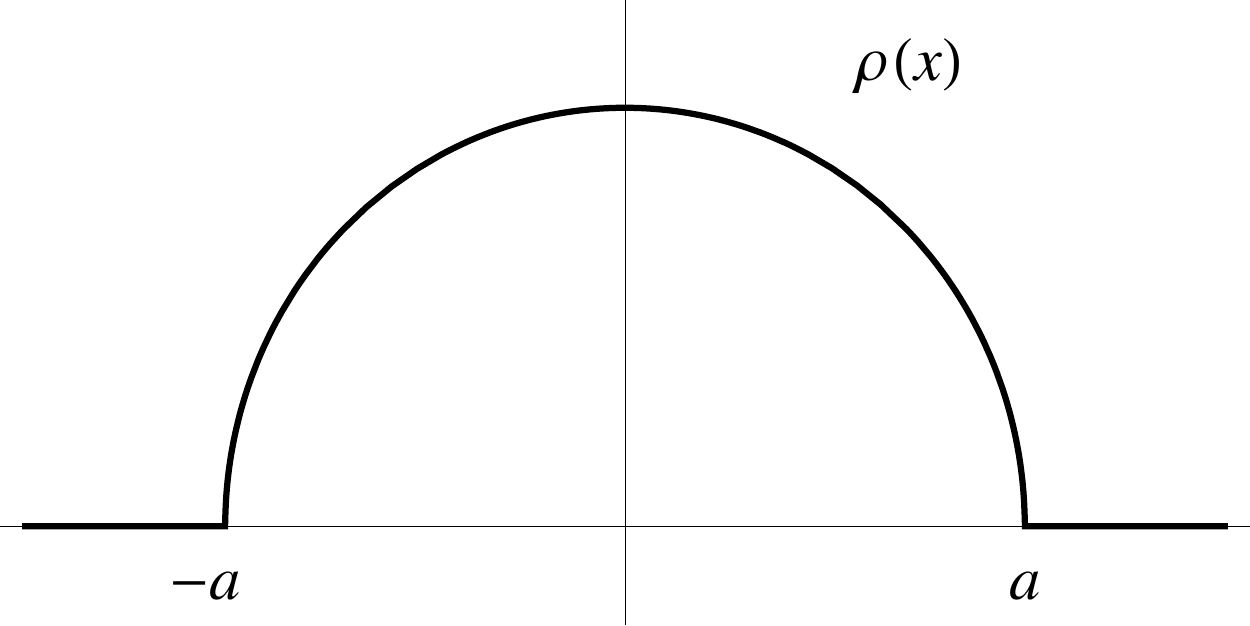}
\caption{Semicircle law of the Gaussian eigenvalue distribution.}
\end{figure}
\end{center}

 \subsection { The Gaussian model as a mother body for elliptic lenses}

The mass distribution of the Gaussian lensing model  of mass $m$
\begin{equation}\label{gaud2}
d \mu_G(x)=\frac{2 m }{\pi a^2}\sqrt{a^2-x^2}\,  d x,\quad -a\leq x\leq a,
\end{equation}
is a  mother body  for $(D,\mu_D)$  \cite{SJ06,SH92b} (see also Appendix A of this paper) where $\mu_D$ is  the uniform mass distribution
   $d \mu_D=d x \,dy$ supported on the elliptic domain
   \begin{equation}\label{app0}
D=\{(x,y)\in \mathbb{R}^2,\, \frac{x^2}{\alpha^2}+\frac{y^2}{\beta^2}< 1,\,\alpha>\beta,\beta>0\},
\end{equation}
 with focal  distance $a=\sqrt{\alpha^2-\beta^2}$  and total mass $m=\pi \alpha \beta$.

 \subsection { The Gaussian model as an edge-on elliptic lens}

According to the discussion of Subsection 2.5,  the  measure $\mu_G$ of the Gaussian model (\ref{gaud2})  describes  an edge-on disk galaxy with uniform mass distribution   of total mass $m$.  The corresponding  domain $\mathcal{D}$ in the $XY$-plane  orthogonal to the lens plane is  bounded by the closed curve with equation
\begin{equation}\label{proal}
Y^2=\frac{S^2}{\pi^2 a^4}\,(a^2-X^2) ,\quad -a\leq X\leq a,
\end{equation}
 where $S$ is the area of $\mathcal{D}$.  Thus $\mathcal{D}$ is an elliptic domain
\begin{equation}\label{app}
\mathcal{D}=\{(X,Y)\in \mathbb{R}^2,\, \frac{X^2}{a^2}+\frac{Y^2}{b^2}< 1\},
\end{equation}
with $b=S/\pi a$.

It is illustrative to provide an alternative direct proof of this interpretation of the Gaussian model  starting from a well-known lensing model \cite{FA09}.  To this end we consider  a uniform mass distribution of total mass $m$ on an elliptic domain in the lens $xy$-plane
\begin{equation}\label{app1}
D=\{(x,y)\in \mathbb{R}^2,\, \frac{x^2}{a^2}+\frac{b^2}{b^2}< 1,\,a>b,b>0\}.
\end{equation}
The corresponding lens equation is \cite{FA09}
\begin{equation}\label{lenapp}
\overline{w}=\overline{z}-\lambda \,\omega(z),\quad \lambda=\frac{m}{\pi a \, b},
\end{equation}
where $\omega(z)$ is given by the expression  (\ref{swap2}) in Appendix A with $(\alpha,\beta)=(a,b)$,  so that
\begin{equation}\label{gaueap}
\left\{\begin{array}{ll} \displaystyle  \overline{w}=\overline{z}-\pi \lambda\Big(\overline{z}-  \frac{(a-b)^2}{c^2} z\Big), \quad z\in D,
 \\\\ \displaystyle
 \overline{w}=\overline{z}-\frac{2m}{c^2}\Big(z-\sqrt{z^2-c^2}\Big), \quad z\in \mathbb{C}\setminus \overline{D},\end{array}\right.
\end{equation}
with $c^2=a^2-b^2$. If we  perform a rotation  of the ellipse plane of angle $\theta$ about the $x$-axis,  the projected mass distribution on the lens plane is now supported on the elliptic domain

\begin{equation}\label{appp}
D'=\{(x,y)\in \mathbb{R}^2,\, \frac{x^2}{a^2}+\frac{y^2}{b'^2}< 1,\,b'=b\cos\theta\},
\end{equation}
and the lens equation reads
 \begin{equation}\label{gaueapi}
\left\{\begin{array}{ll} \displaystyle  \overline{w}=\overline{z}-\pi \lambda'\Big(\overline{z}-  \frac{(a-b')^2}{c'^2} z\Big), \quad z\in D',
 \\\\ \displaystyle
 \overline{w}=\overline{z}-\frac{2m}{c'^2}\Big(z-\sqrt{z^2-c'^2}\Big), \quad z\in \mathbb{C}\setminus \overline{D'},\end{array}\right.
\end{equation}
where
\begin{equation}\label{par}
c'^2=a^2-b'^2,\quad  \lambda'=\frac{m}{\pi a \, b'}.
\end{equation}
 If the inclination angle $\theta$  tends to $\pi/2$, then we are led to an edge-on galaxy with
 \begin{equation}\label{hhh}
 a={\rm Constant},\quad b'\rightarrow 0.
 \end{equation}
 Thus $D'$ shrinks to the interval $[-a,a]$ and
 \begin{equation}\label{lims}
 c'\rightarrow a,\quad \pi \lambda'\Big( 1-\frac{(a-b')^2}{c'^2}\Big)\rightarrow \frac{2m}{a^2}.
 \end{equation}
 Therefore, the lens equation (\ref{gaueapi}) becomes exactly the lens equation (\ref{gaue1}) provided by the Gaussian model  of mass $m$.

\subsection{ Solutions of the lens equation}

From the first equation of (\ref{gaue1}) we have that there is a unique  dim image $x=w/(1-p)$
for any
source $w$ in the interval
\begin{equation}\label{isu2}
-a\Big|1-p \Big |\leq w\leq a\Big|1-p \Big |,
\end{equation}
and no dim image otherwise. It should be noticed that for $p=1$ the whole interval $[-a,a]$ becomes the image of the origin $w=0$.

Let us now consider the lens equation (\ref{gaue1}) for bright images. The presence of the term with a square root  suggests the introduction of a Joukowski change  of variable
\begin{equation}\label{juk}
z=\frac{a}{2}\Big(Z+\frac{1}{Z} \Big),
\end{equation}
which defines a conformal one-to-one map of  the domain $\mathbb{C}\setminus [-a,a]$ in the $z$-plane onto the domain $|Z|<1$ in the $Z$-plane
\begin{equation}\label{juk1}
Z=\frac{a}{z+\sqrt{z^2-a^2}}.
\end{equation}
Thus  the lens equation (\ref{gaue1}) for bright images   becomes
\begin{equation}\label{qau2}
 Z^2-2 p |Z|^2-2 u Z+1=0 , \quad |Z|<1,
\end{equation}
where
\begin{equation}\label{nsu}
u:=\frac{w}{a},
\end{equation}
denotes the normalized source position.
Then if we set
\begin{equation}\label{comp}
Z=\mathrm{x}+i \mathrm{y},\quad u=\alpha+i \beta,
\end{equation}
we have that the solutions of (\ref{qau2}) are  the intersection points  of the  pair of conics
\begin{equation}\label{sis}
\left\{\begin{array}{ll} \beta \mathrm{x} +\alpha \mathrm{y}-\mathrm{x} \mathrm{y}=0,\\\\   (1-2p)\, \mathrm{x} ^2-(1+2p)\, \mathrm{y}^2-2(\alpha \mathrm{x} -\beta \mathrm{y})+1=0, \end{array}\right.
\end{equation}
which satisfy
 \begin{equation}\label{dei}
 \mathrm{x} ^2+\mathrm{y}^2<1.
 \end{equation}
 Hence, we deduce that there are at most four bright images. Since $\mu_G$ is a mother body for a uniform distribution of an elliptic domain, this result is in agreement with Theorem 5.1 of \cite{FA09}. We also note that the system (\ref{sis}) is invariant under the transformations $(\alpha, \mathrm{x})\leftrightarrow (-\alpha,-\mathrm{x})$ and $(\beta,\mathrm{y})\leftrightarrow(-\beta,-\mathrm{y})$ so that we may restrict our analysis to the first quadrant $\alpha\geq0, \beta\geq0$ of the $u$-plane.

If we eliminate $\mathrm{y} $ as $\mathrm{y} =\beta \mathrm{x}/(\mathrm{x}-\alpha)$ from the first equation of (\ref{sis}) and substitute the result  into the second equation of (\ref{sis}),  we get the following quartic equation for $\mathrm{x}$
\begin{eqnarray}\label{eqy}
\nonumber && (1-2p)\mathrm{x}^4+4\alpha (p-1)\mathrm{x}^3+(5\alpha^2+\beta^2-2p(\alpha^2+\beta^2)+1) \mathrm{x}^2 \\
&&-2\alpha (\alpha^2+\beta^2+1)\mathrm{x}+\alpha^2=0.
\end{eqnarray}
Then the solutions of (\ref{sis}) can be written as algebraic expressions in terms of the coefficients of (\ref{eqy}).

  For example if $u=0$ and $p>1$  there are  four  distinct  solutions of (\ref{sis})-(\ref{dei}) (see Fig. 4): two purely imaginary
\begin{equation}\label{see}
Z^{(1)}_{\pm}=\pm \frac{i}{\sqrt{1+2 p}},
\end{equation}
which  determine two purely imaginary images in the $z$-plane
\begin{equation}\label{seez}
z^{(1)}_{\pm}=\mp i\frac{ap}{\sqrt{1+2 p}},
\end{equation}
and two real solutions
\begin{equation}\label{see2}
Z^{(2)}_{\pm}=\pm \frac{1}{\sqrt{2 p-1}},
\end{equation}
which  determine two real images in the $z$-plane
\begin{equation}\label{see2z}
z^{(2)}_{\pm}=\pm \frac{ap}{\sqrt{2 p-1}}.
\end{equation}

\begin{center}
\begin{figure}[ht]
\centering
\includegraphics[width=7cm]{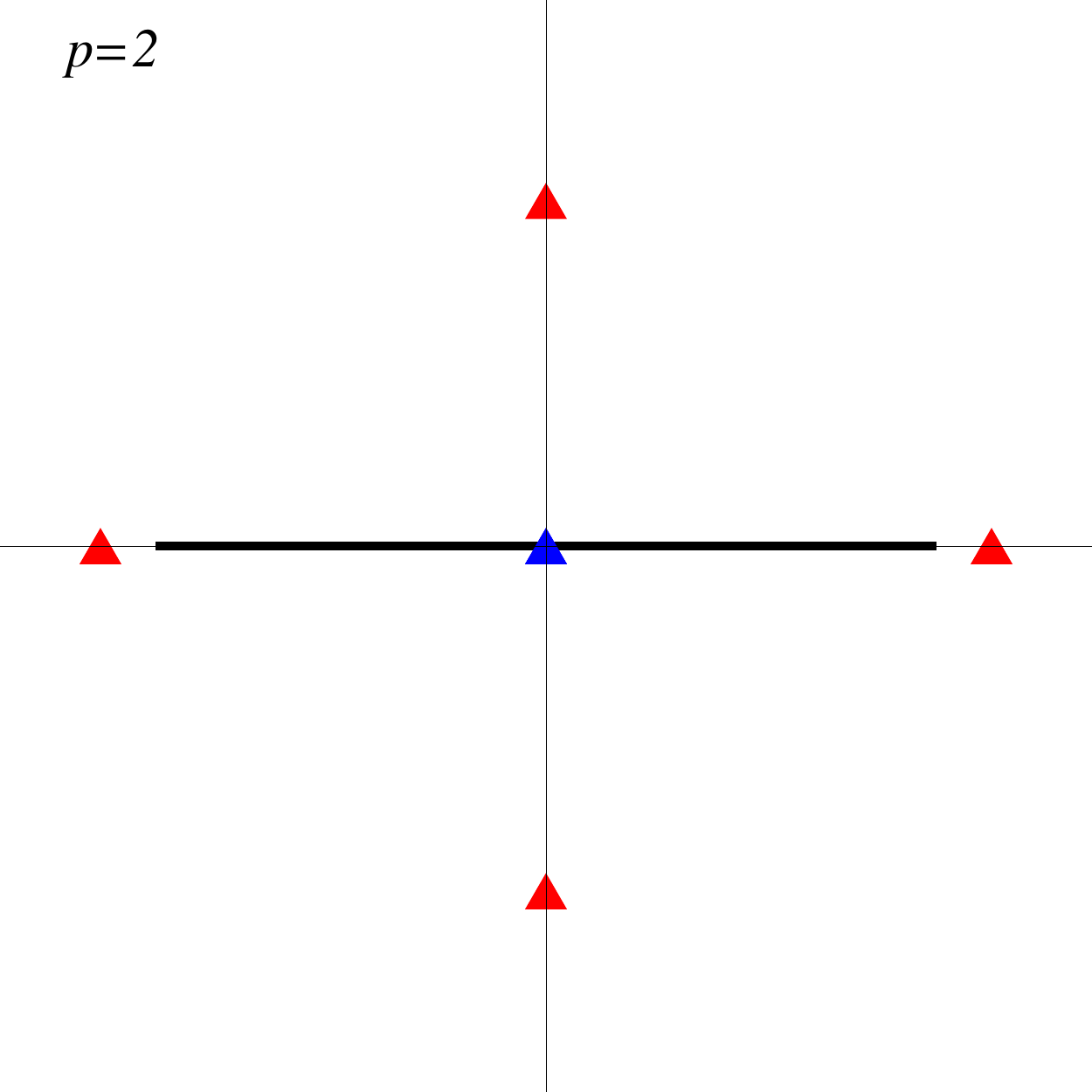}
\caption{ The Gaussian model for $p=2$ exhibits  an Einstein cross of  images of the source $w=0$, with four bright images (red triangles) and one dim images (blue triangle).}
\end{figure}
\end{center}

\subsection{The case  $p=1/2$}

 For $p=1/2$  the equation (\ref{eqy}) reduces to a cubic  equation and it is easy to complete the explicit analysis of the images. 
 
 The bright images for  on-axis  sources are as follows:
\begin{enumerate}
\item For  $\alpha=\beta=0$
there are two solutions of (\ref{sis}) which satisfy (\ref{dei}) given by $Z_{\pm}=\pm i/\sqrt{2}$. They correspond to $z_{\pm}=\mp i a/(2\sqrt{2})$ in the $z$-plane.
\item  For  $\beta=0,\,\alpha\,>0$
there are three  solutions of (\ref{sis})
\begin{equation}\label{Zs}
Z_{\pm}=\alpha\pm i\sqrt{\frac{1}{2}-\alpha^2},\qquad Z_0=\frac{1}{2\alpha}.
\end{equation}
They satisfy (\ref{dei}) for the following values of $\alpha$
\begin{equation}\label{axesalpha}
\left\{\begin{array}{ll}
Z_{\pm} & \mbox{if }\quad 0<\alpha\leq\frac{1}{2},\\
Z_{\pm},\, Z_0 & \mbox{if }\quad \frac{1}{2}<\alpha<\frac{1}{\sqrt{2}},\\
Z_0 & \mbox{if }\quad \alpha\geq\frac{1}{\sqrt{2}}.
\end{array}\right.
\end{equation}
The corresponding solutions in the $z$-plane are
\begin{equation}\label{Zs1}
z_{\pm}=\frac{a}{2}\Big( 3\alpha\mp i\sqrt{\frac{1}{2}-\alpha^2}\Big),\qquad z_0=a\Big(\alpha+\frac{1}{4\alpha}\Big).
\end{equation}
\item  For  $\alpha=0\,, \beta\,>0$
the system (\ref{sis}) has two solutions
\begin{equation}\label{hat1}
\widehat{Z}_{\pm}=\frac{i}{2}\left(\beta\pm \sqrt{\beta^2+2}\right),
\end{equation}
which satisfy  (\ref{dei}) for the following values of $\beta$
\begin{equation}\label{axesbeta}
\left\{\begin{array}{ll}
\widehat{Z}_{\pm} & \mbox{if }\quad 0<\beta<\frac{1}{2},\\
\widehat{Z}_-  & \mbox{if }\quad \beta\geq\frac{1}{2}.
\end{array}\right.
\end{equation}
The corresponding solutions in the $z$-plane are
\begin{equation}\label{hatc}
\widehat{z}_{\pm}=i\,\frac{a}{4}\left(3\beta\mp \sqrt{\beta^2+2}\right).
\end{equation}

\end{enumerate}

 The images for  off-axis sources with 
\begin{equation}\label{out}
\alpha,\,\beta\,>0,
\end{equation}
 can be determined as follows. Since  $\alpha$ and $\beta$ are different from zero,  we have that $\mathrm{x}\neq \alpha$ and $\mathrm{y} =\beta \mathrm{x}/(\mathrm{x}-\alpha)$. Hence (\ref{eqy}) reduces to the cubic equation
 \begin{equation}\label{sisq}
 -2 \alpha \mathrm{x}^3+(4\alpha^2+1) \mathrm{x}^2-2\alpha (\alpha^2+\beta^2+1)\mathrm{x}+\alpha^2=0.
\end{equation}
 The discriminant of the polynomial in this equation vanishes on the curve
 \begin{equation}\label{disc}
16\alpha^6+8\alpha^4(4\beta^2-3)+4\alpha^2(4\beta^4+10\beta^2+3)-(\beta^2+2)=0.
\end{equation}
Then, it is straightforward to deduce that
\begin{enumerate}
\item For $u$ outside the curve (\ref{disc}) there are one or  three different solutions of (\ref{sisq}).
\item For $u$ on the curve (\ref{disc}) there are  two different solutions of (\ref{sisq}).
\end{enumerate}

Finally we have to determine which solutions of (\ref{sisq}) satisfy the condition (\ref{dei}). Now
from (\ref{sis})
it follows that $M= \mathrm{x} ^2+\mathrm{y}^2$  satisfies the equation
\begin{equation}\label{m}
16\alpha^2M^3-4(1+4\alpha^2)M^2+4(1+\alpha^2+\beta^2)M-1=0.
\end{equation}
Then from Bolzano's theorem it is immediate to deduce that one of the solutions of (\ref{m}) satisfies $M<\frac{1}{2}$ . Moreover, for $|u|<1/2$ we have that at least
one solution  of (\ref{m})  satisfies $\frac{1}{2}<M<1$, and for $|u|=1/2$ there exists a solution with $M=1$.

Therefore, we conclude that the number of bright images  depends on the  relative position $u$ of the source as shown in Fig.  5.

\begin{center}
\begin{figure}[ht]
\centering
\includegraphics[width=5cm]{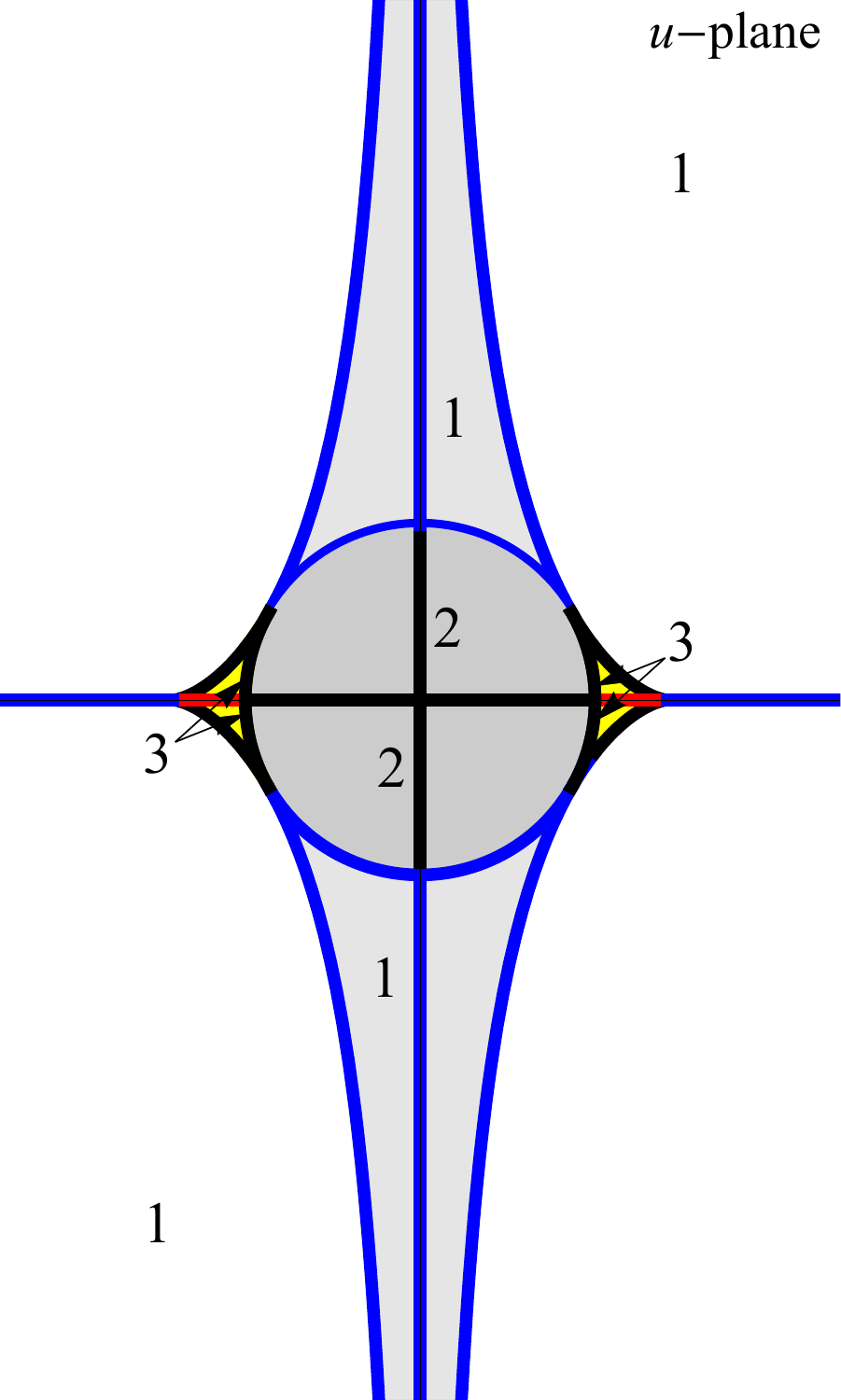}
\caption{Number of bright images  for the different  positions of the source in the $u$-plane for the case  $p=\frac{1}{2}$. Red, black and blue colors on the axes and on the curve stand for three, two and one bright images respectively.  The different  configurations are determined by the relative position of the source with respect to the curve (\ref{disc}) and the disk $|u|<1/2$.}
\end{figure}
\end{center}

 \subsection{ Time delays }

Using the expression (\ref{gaud}) of the mass density we obtain the explicit form of the logarithmic potential 
\begin{equation}\label{logp}
\fl -\int_{-a}^a \rho(x)\ln |z-x|\, d x=-\frac{1}{a^2}\,{\rm Re}(z^2-z\sqrt{z^2-a^2})-\ln|z+\sqrt{z^2-a^2}|+\frac{1}{2}+\ln 2.
\end{equation}
Hence, from (\ref{arr}) we may determine the time delay for bright  images. In particular, for the bright images (\ref{seez}) and (\ref{see2z}) of  $w=0$ we have
\begin{equation}\label{arrr}
\tau(z^{(1)}_{\pm})=\frac{m}{2}\Big(1+ \ln \frac{1}{m+(a/2)^2
}\Big),
\end{equation}
\begin{equation}\label{arrr2}
\tau(z^{(2)}_{\pm})=\frac{m}{2}\Big(1+ \ln \frac{1}{m-(a/2)^2}\Big),
\end{equation}
so that the relative  time delay for the reception of both pairs of images is given by
\begin{equation}\label{arrr3}
\tau(z^{(1)}_{\pm})-\tau(z^{(2)}_{\pm})=\frac{m}{2}\, \ln \Big(\frac{m-(a/2)^2}{m+(a/2)^2}\Big).
\end{equation}

\section{The quartic  model \label{sec:quar}}

We consider the quartic matrix model  defined by the potential function
\begin{equation}\label{qua}
V(z)=\frac{z^4}{4}+t\,\frac{z^2}{2},
\end{equation}
where $t$ is a real parameter. The corresponding eigenvalue distribution has a  one-cut  support    for $ t> -\sqrt{2}$ and a two-cut support for $ t< -\sqrt{2}$  \cite{BL08} (see also \cite{BL03}). The value $t=-\sqrt{2}$ represents a point of phase transition.

\subsection{The one-cut distribution}
For $ t>-\sqrt{2}$ the
function $P(z)$ in (\ref{alg}) is given by
\begin{equation}\label{qee2}
P(z)= (z^2-a^2)(z^2+c)^2,
\end{equation}
where
\begin{equation}\label{dens}
a=\sqrt{\frac{2}{3}}\sqrt{-t+\sqrt{t^2+6} }, \quad c=\frac{1}{3}\Big(2t+\sqrt{t^2+6} \Big) .
\end{equation}
The associated mass density is (Fig. 6)
\begin{equation}\label{qud1}
 \rho(x)=\frac{1}{\pi}(x^2+c)\sqrt{a^2-x^2},\quad -a\leq x\leq a.
\end{equation}
\begin{center}
\begin{figure}[ht]
\centering
\includegraphics[width=9cm]{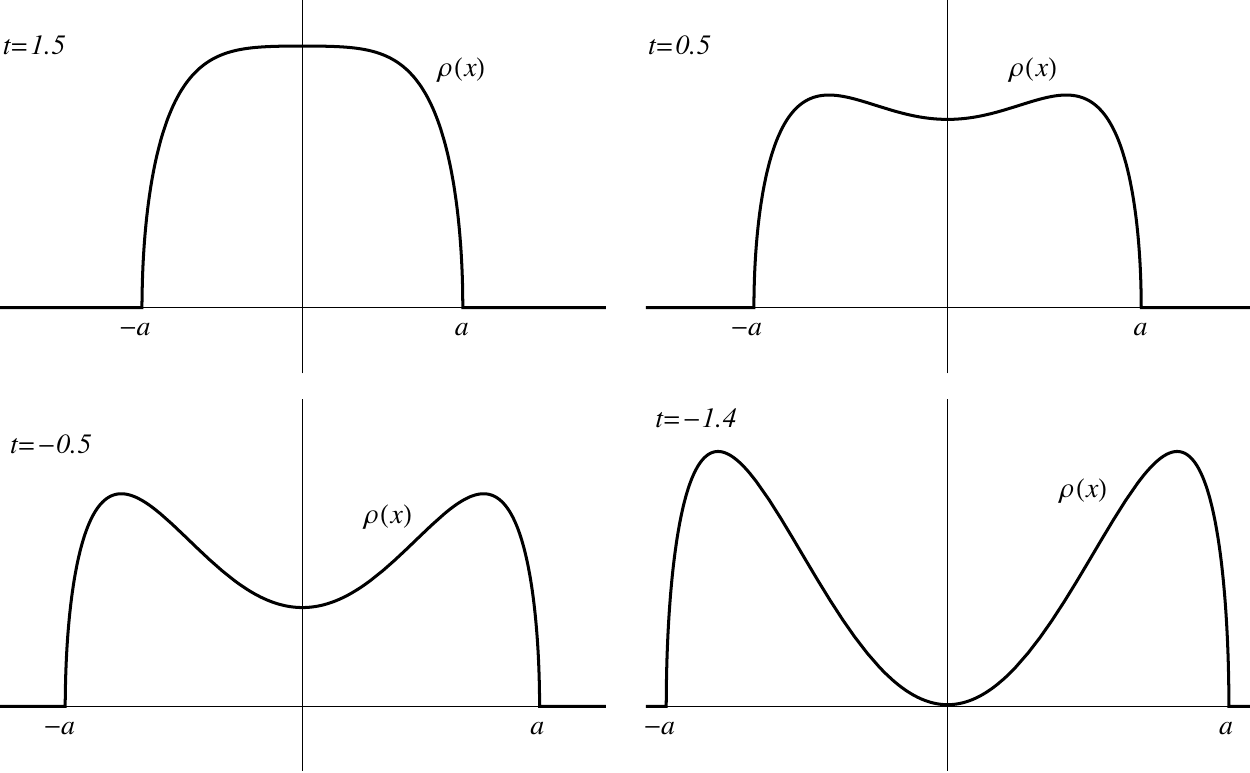}
\caption{Eigenvalue density of the quartic model  for several values of $t$ in the one-cut case.}
\end{figure}
\end{center}

 \subsection { The one-cut  distribution as a mother body for elliptic lenses}

As it is shown in the appendix A of this paper,  the measure $d \mu_Q(x)=m\rho(x) d x$ determined by  the density (\ref{qud1})  is a  mother body  for the planar body $(D,\mu_D)$ where
$D$ is any elliptic domain $D$  (\ref{app0}) contained in the region
\begin{equation}\label{posi}
y^2<x^2+\frac{c_3}{2 c_2},
\end{equation}
with
\begin{equation}\label{solui}
 c_2=\frac{3}{3 A_1^2+A_2^2+3},\quad c_3=c+\frac{a^2 A_2^2}{3 A_1^2+A_2^2+3},
\end{equation}
\begin{equation}\label{aesi}
A_1=\frac{\alpha^2+\beta^2}{a^2},\quad A_2=-2\frac{\alpha \beta}{a^2},\quad a^2=\alpha^2-\beta^2,
\end{equation}
and $\mu_D$ is the measure supported  on $D$ given by
\begin{equation}\label{posi1}
d \mu_D(x,y)=\frac{m}{\pi |A_2|}\Big(2 c_2(x^2-y^2)+c_3\Big) d x dy.
\end{equation}

\subsection { The one-cut distribution as an edge-on  lens}

According to  (\ref{proa}) the one-cut distribution  $d \mu_Q(x)=m\rho(x) d x$  of the quartic model describes an  edge-on  lens with a uniform mass distribution of total mass $m$
 on the region $\mathcal{D}$ of the $XY$-plane bounded by the curve
\begin{equation}\label{proa1}
Y^2=\frac{S^2}{4\pi^2}(X^2+c)^2(a^2-X^2)  ,\quad -a<X<a,
\end{equation}
where $S$ is the area of $\mathcal{D}$ (Fig. 7).

\begin{center}
\begin{figure}[ht]
\centering
\includegraphics[width=9cm]{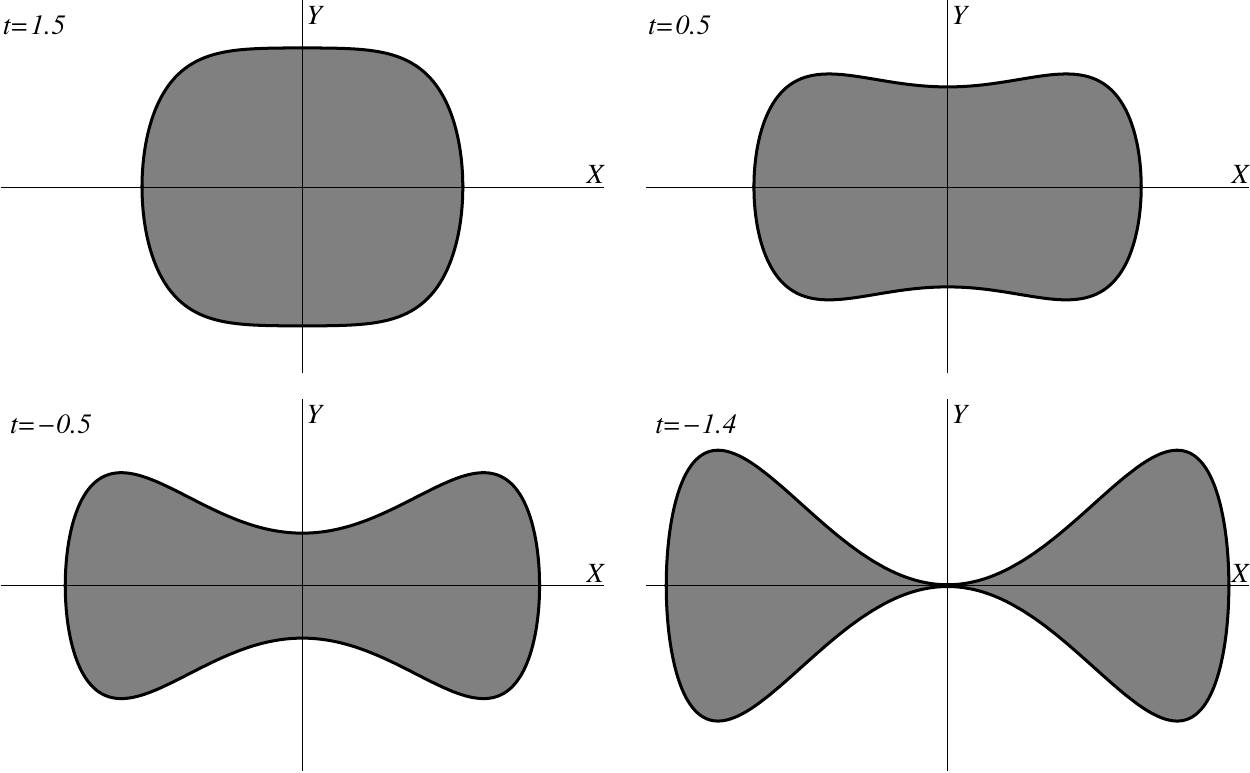}
\caption{Edge-on galaxies for the quartic model in the one-cut case.}
\end{figure}
\end{center}

\subsection{ Solutions of the lens equation in the one-cut case}

The lens equation in the one-cut case  takes the form
\begin{equation}\label{qua1}
\left\{\begin{array}{ll}  w=x-m (x^3+t x) ,\quad  -a\leq x\leq a, \\\\   \overline{w}=\overline{z}-m\Big(z^3+tz-(z^2+c)\sqrt{z^2-a^2}\Big)  , \quad z\in \mathbb{C}\setminus [-a,a].\end{array}\right.
\end{equation}
We will concentrate on the images produced by the source $w=0$.

The dim images for $w=0$  are the solutions of
\begin{equation}\label{w0donecut}
x\left[m x^2+(m t-1)\right]=0,\quad x\in[-a,a].
\end{equation}
This equation has always the solution  $x^{(0)}=0$. Furthermore, if one of the  two pairs of conditions
\begin{equation}\label{con1d}
\begin{array}{l}
m>\frac{1}{\sqrt{2}},\quad t\in[-\sqrt{2},\frac{1}{m}),\\
m\in(0,\frac{1}{\sqrt{2}}],\quad t\in[t_c,\frac{1}{m}),
\end{array}
\end{equation}
is satisfied, where
\begin{equation}\label{tec}
t_c=\frac{2\sqrt{1-2m^2}-1}{m},
\end{equation}
 then there is a  pair of additional images
\begin{equation}\label{dx}
x^{(d)}_{\pm}=\pm\sqrt{\frac{1}{m}-t}.
\end{equation}

The bright images of $w=0$ are characterized by the equation
\begin{equation}\label{w0bonecut}
m(z^3+t z)-\overline{z}=m(z^2+c)\sqrt{z-a}\sqrt{z+a},\quad z\in{\mathbb C}\setminus[-a,a],
\end{equation}
where we assume the principal branch for  both square roots.
 It is easy to check   that (\ref{w0bonecut}) has not off-axis solutions. 
The bright images  $x$ on the real axis  satisfy
\begin{equation}\label{re1b}
\left\{\begin{array}{ll}
m x^3+(m t-1)x=m(x^2+c)\sqrt{x^2-a^2}, & \mbox{if }\;x>a,\\
m x^3+(m t-1)x=-m(x^2+c)\sqrt{x^2-a^2}, & \mbox{if }\;x<-a.
\end{array}\right.
\end{equation}
Then there exist  two solutions $x^{(1)}_{\pm}$  $(x^{(1)}_-=-x^{(1)}_+)$   if and only if one of the following
three  pairs of conditions is verified :
\begin{equation}\label{conre1b}
\begin{array}{l}
m>\frac{1}{\sqrt{2}},\quad t\in[-\sqrt{2},\infty),\\
m=\frac{1}{\sqrt{2}},\quad t\in(-\sqrt{2},\infty),\\
m\in(0,\frac{1}{\sqrt{2}}),\quad t\in(t_c,\infty).
\end{array}
\end{equation}
 Note that $t_c>-\sqrt{2}$ if $m<\frac{1}{\sqrt{2}}$.

 The bright images $i\,y \, (y\neq 0)$ on the imaginary axis are the solutions of
\begin{equation}\label{im1b}
\left\{ \begin{array}{ll}
-m y^3+(m t+1)y=m(c-y^2)\sqrt{y^2+a^2},& \mbox{if }\,y>0,\\
-m y^3+(m t+1)y=-m(c-y^2)\sqrt{y^2+a^2},& \mbox{if }\,y<0.
\end{array}\right.
\end{equation}
Then, it follows that  there are two solutions $i y^{(1)}_{\pm}$  $(y^{(1)}_-=-y^{(1)}_+)$  if and only if one of the following
two pairs of  conditions is verified:
\begin{equation}\label{conim1b}
\begin{array}{l}
m>\frac{1}{\sqrt{2}},\quad t\in[-\sqrt{2},\infty ),\\
m\in(0,\frac{1}{\sqrt{2}}],\quad t\in(-\sqrt{2},\infty).
\end{array}
\end{equation}
\begin{center}
\begin{figure}[ht]
\centering
\includegraphics[width=11cm]{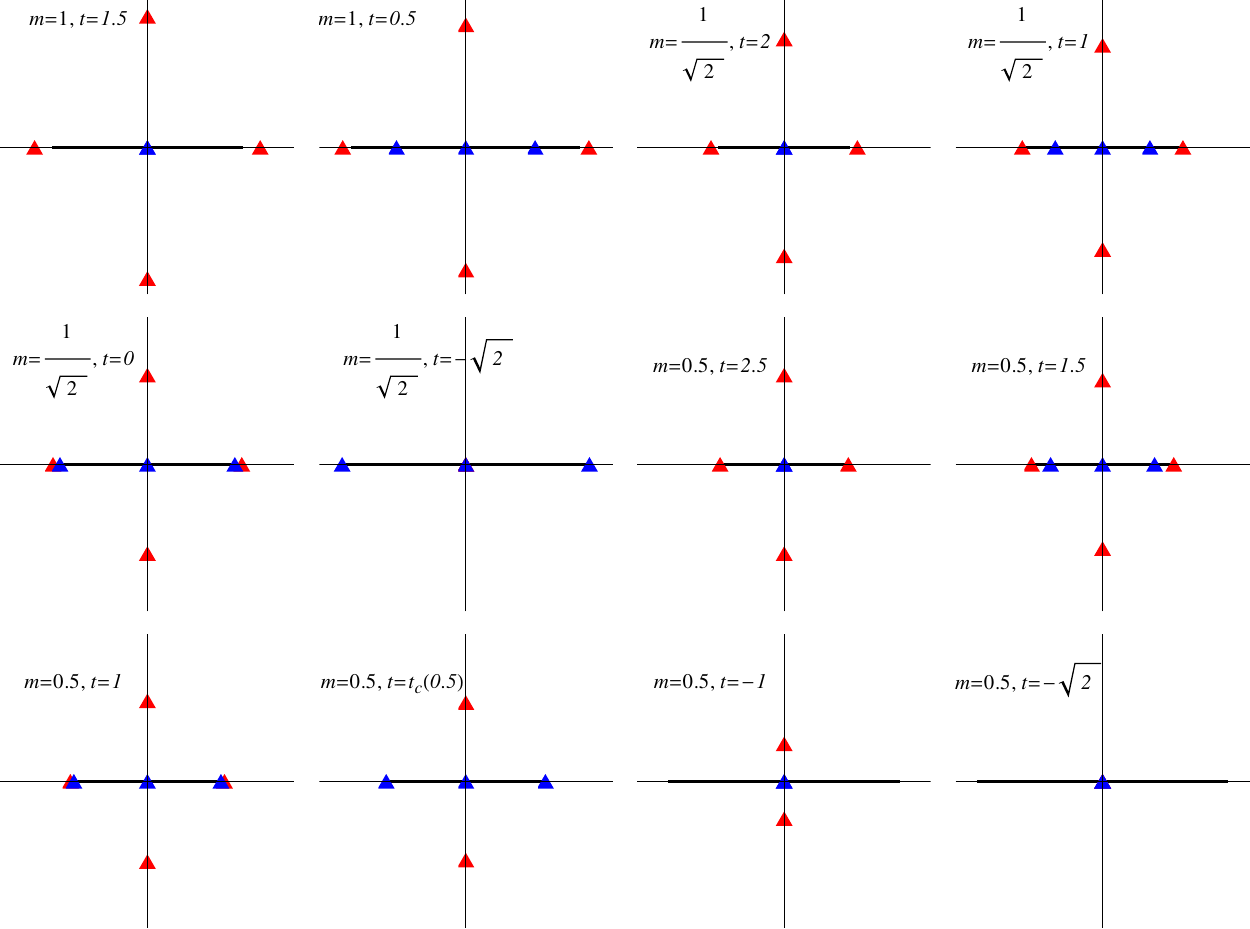}
\caption{Sets of images of $w=0$  in the one cut case. The first two figures of the first row correspond to $m=1>1/\sqrt{2}$ and represent an Einstein cross at $t=1.5>1/m$ which becomes accompanied by a couple of dim images at $t=0.5<1/m$. The next four figures illustrate the case $m= 1/\sqrt{2}$. They show an Einstein cross at $t=2>\sqrt{2}$  and a  couple of dim images which arises for $t<\sqrt{2}$ and move, together with the two bright images on the real axis, to the endpoints of the lens at $t=-\sqrt{2}$. All the bright images dissapear at $t=-\sqrt{2}$. Finally, the last six figures for $m=0.5<1/\sqrt{2}$ show a process  in which two dim images and two bright images move towards the endpoints of the lens. For $t<t_c$ only the dim image at the origin and the two bright images on the imaginary axis appear. These last two images vanish at $t=-\sqrt{2}$.}
\end{figure}
\end{center}
In conclusion,  the  images of $w=0$  in the one-cut case are classified into  the following cases (Fig. 8):
\begin{equation}\label{resonecut}\everymath{\displaystyle}
\begin{array}{lll}
m>\frac{1}{\sqrt{2}}\,:                  & x^{(0)},\,x^{(1)}_{\pm},\;iy^{(1)}_{\pm}                       &\mbox{for}\quad t\geq\frac{1}{m},                       \\
                                         & x^{(0)},\,x^{(d)}_{\pm},\,x^{(1)}_{\pm},\;iy^{(1)}_{\pm}       &\mbox{for}\quad t\in\left[-\sqrt{2},\frac{1}{m}\right), \\  \\
m=\frac{1}{\sqrt{2}}\,:                  &x^{(0)},\, x^{(1)}_{\pm},\, iy^{(1)}_{\pm}                      &\mbox{for}\quad t\geq\sqrt{2},                          \\
                                         &x^{(0)},\;x^{(d)}_{\pm},\,x^{(1)}_{\pm},\;iy^{(1)}_{\pm}        &\mbox{for}\quad t\in(-\sqrt{2},\sqrt{2}),               \\
                                         &x^{(0)},\;x^{(d)}_{\pm}                                         &\mbox{for}\quad t=-\sqrt{2},                            \\  \\
m\in\left(0,\frac{1}{\sqrt{2}}\right)\,: &x^{(0)},\,x^{(1)}_{\pm},\;iy^{(1)}_{\pm}                        &\mbox{for}\quad t\geq\frac{1}{m},                       \\
                                         & x^{(0)},\;x^{(d)}_{\pm},\,x^{(1)}_{\pm},\;iy^{(1)}_{\pm}       &\mbox{for}\quad t\in\left(t_c,\frac{1}{m}\right),       \\
                                         & x^{(0)},\;x^{(d)}_{\pm},\,iy^{(1)}_{\pm}                       &\mbox{for}\quad t=t_c,                                  \\
                                         & x^{(0)},\,iy^{(1)}_{\pm}                                       &\mbox{for}\quad t\in(-\sqrt{2},t_c),                    \\
                                         & x^{(0)}                                                        &\mbox{for}\quad t=-\sqrt{2}.
\end{array}
\end{equation}

\subsection{ The two-cut distribution}

 For $t<-\sqrt{2}$ the quartic matrix model exhibits a  two-cut  distribution (\cite{BL08}, \cite{BL03}) such that the function $P(z)$ in (\ref{alg}) is
 \begin{equation}\label{qee1}
P(z)=z^2(z^2-a^2)(z^2-b^2),
\end{equation}
 where
 \begin{equation}\label{end}
 a=\sqrt{\sqrt{2}-t},\quad b=\sqrt{-\sqrt{2}-t}.
 \end{equation}
It leads to a mass density (Fig. 9):
\begin{equation}\label{qud}
 \rho(x)=\frac{|x|}{\pi}\sqrt{(a^2-x^2)(x^2-b^2)},\quad  x\in [-a,-b] \cup[b,a],
 \end{equation}
 and to a Cauchy transform
 \begin{equation}\label{ccc}
 \omega(z)=z^3+tz-z\sqrt{(z^2-a^2)(z^2-b^2)}.
 \end{equation}
  \begin{center}
\begin{figure}[ht]
\centering
\includegraphics[width=10cm]{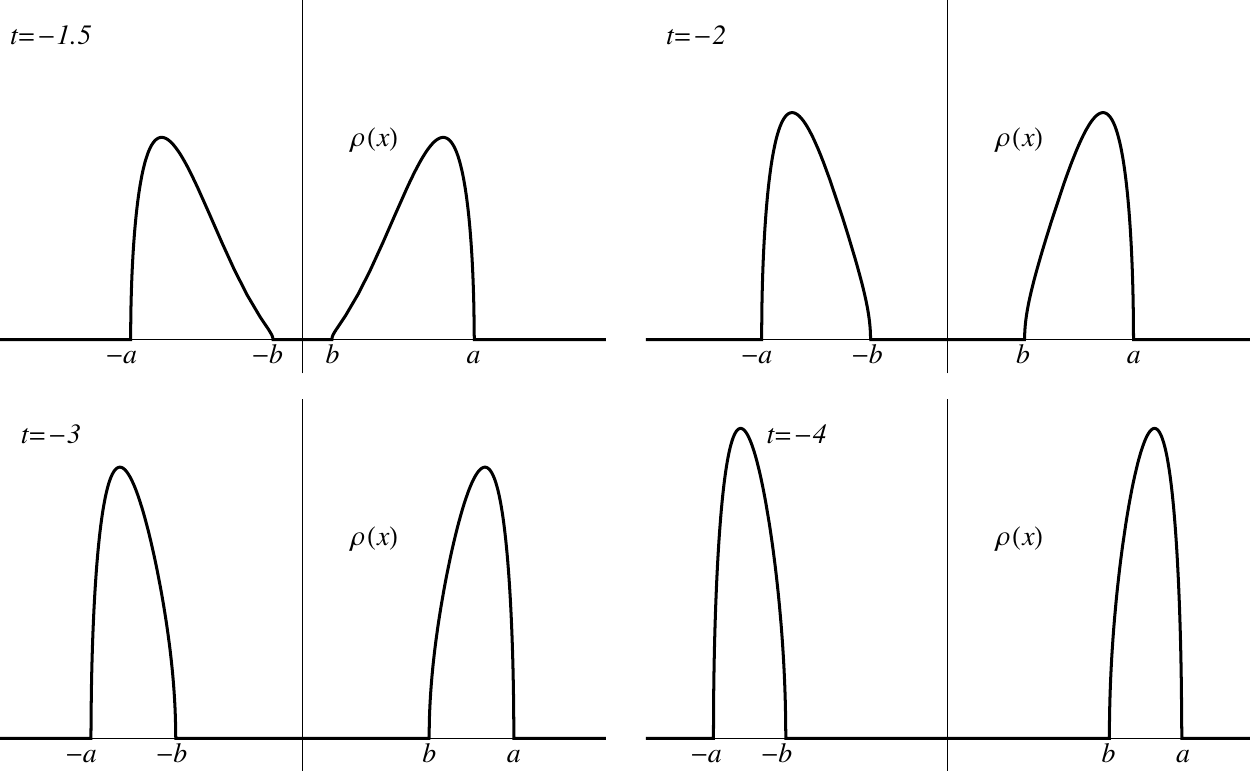}
\caption{Eigenvalue densities of the quartic model in the two-cut case.}
\end{figure}
\end{center}

 \subsection { The two-cut  distribution as an edge-on double lens}

According to (\ref{proa}) the two cut distribution of the quartic model describes an edge-on  lens composed by two twin galaxies  with a uniform mass distribution of total mass $m$
 supported on the regions $\mathcal{D}_{\pm}$ of the $XY$-plane bounded by the two-component curve
\begin{equation}\label{proa2}
Y^2=\frac{S^2}{4\pi^2}X^2(a^2-X^2)(X^2-b^2)  ,\quad x\in [-a,-b] \cup[b,a] ,
\end{equation}
where $S$ is the area of  $\mathcal{D}_-\cup \mathcal{D}_+$ (Fig. 10).

\begin{center}
\begin{figure}[ht]
\centering
\includegraphics[width=10cm]{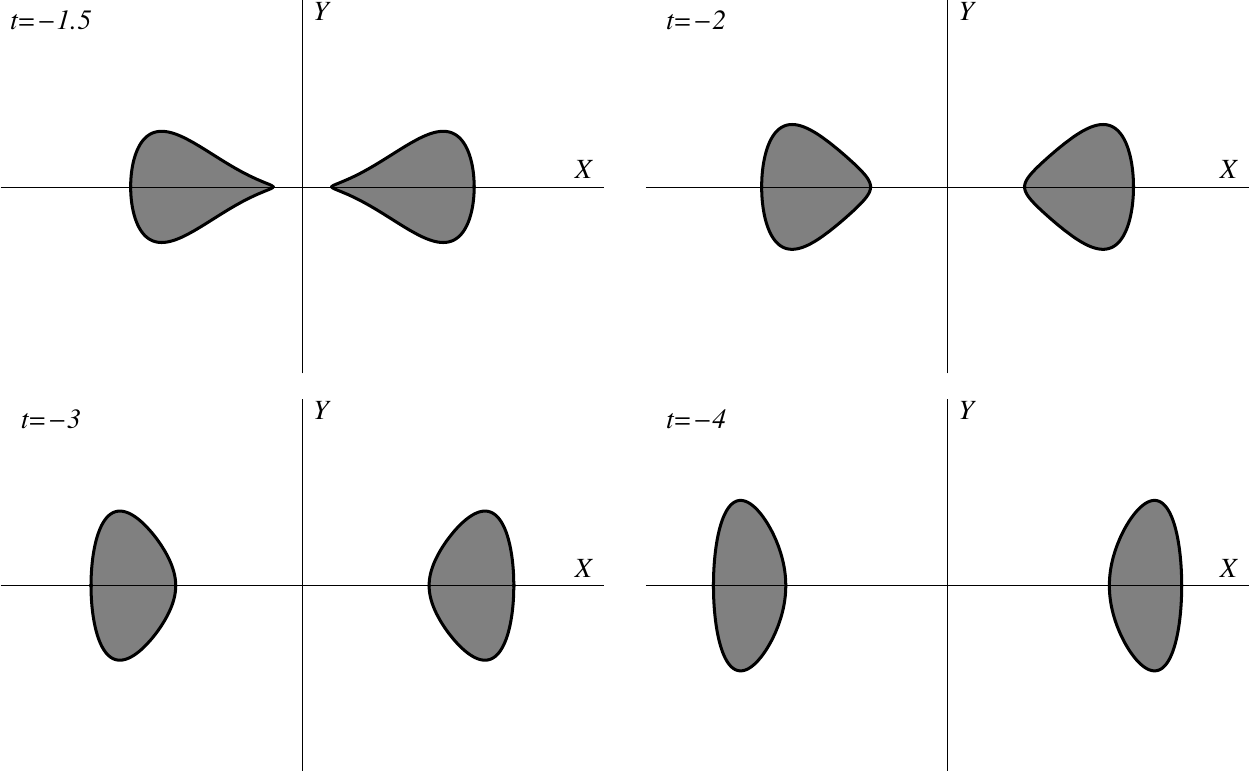}
\caption{Edge on galaxies for the quartic model in the two cut case.}
\end{figure}
\end{center}

\subsection{ Solutions of the lens equation in the  two-cut case}

The lens equation for the two-cut distribution  is

\begin{equation}\label{quar2}
\fl \left\{\begin{array}{ll}  w=x-m (x^3+t x) ,\quad  x\in [-a,-b] \cup[b,a], \\\\   \overline{w}=\overline{z}-m\Big(z^3+tz-z\sqrt{(z^2-a^2)(z^2-b^2)}\Big)  , \quad z\in \mathbb{C}\setminus \Big([-a,-b] \cup[b,a]\Big).\end{array}\right.
\end{equation}
We again concentrate on the images of the source  $w=0$.

The dim images are the solutions of the
equation
\begin{equation}\label{w0dtwocut}
m(x^2+t)-1=0,\quad x\in[-a,-b]\cup[b,a].
\end{equation}
Then if $m\geq\frac{1}{\sqrt{2}}$ we have two dim images at the positions
$$x_{\pm}^{(d)}=\pm\sqrt{\frac{1}{m}-t},$$
and no dim images otherwise.

The  bright images of $w=0$ satisfy
\begin{equation}\label{w0btwocut}
\fl \overline{z}-m\Big(z^3+tz-z\sqrt{(z^2-a^2)(z^2-b^2)}\Big) =0,\quad z\in{\mathbb C}\setminus\Big([-a,-b]\cup[b,a]\Big).
\end{equation}
 Again, a direct computation shows  that (\ref{w0btwocut})  has not
off-axis solutions.

The real solution $x^{(0)}=0$ is now a bright image and  from (\ref{end}) it can be easily seen that  other real solutions $x$
must satisfy
\begin{equation}\label{re2b}
\left\{\begin{array}{ll}
m x^2+(m t-1)=m\sqrt{(x^2+t)^2-2}, & \mbox{if }\;|x|>a,\\
m x^2+(m t-1)=-m\sqrt{(x^2+t)^2-2}, & \mbox{if }\;|x|<b.
\end{array}\right.
\end{equation}
It turns out that  there are not solutions such that $|x|<b$. Nevertheless,  for $m>\frac{1}{\sqrt{2}}$ the solutions  $|x|>a$ are given by
$$x_{\pm}^{(2)}=\pm\sqrt{m+\frac{1}{2m}-t}.$$

The bright images  $z=i y$, $(y\neq0)$  on the imaginary axis  satisfy
\begin{equation}\label{im2b}
m y^2-(m t+1)=m \sqrt{(y^2-t)^2-2}.
\end{equation}
Then we get  the solutions
\begin{equation}\label{imas}
i y_{\pm}^{(2)}=\pm i\sqrt{m+\frac{1}{2m}+t},
\end{equation}
provided  $m>1/\sqrt{2}$ and $t\in\left(-m-\frac{1}{2m},-\sqrt{2}\right).$

Summarizing, the dim and bright images of $w=0$ in the two-cut case are
\begin{equation}\label{restwocut}\everymath{\displaystyle}
\begin{array}{lll}
m>\frac{1}{\sqrt{2}}\,:                  & x^{(0)},\,x^{(d)}_{\pm},\,x^{(2)}_{\pm},\,i y^{(2)}_{\pm}     & \mbox{for}\quad t\in\left(-m-\frac{1}{2m},-\sqrt{2}\right), \\
                                         & x^{(0)},\,x^{(d)}_{\pm},\,x^{(2)}_{\pm},                      & \mbox{for}\quad t\leq-m-\frac{1}{2m},\\  \\
m=\frac{1}{\sqrt{2}}\,:                  & x^{(0)},\;x^{(d)}_{\pm}=\pm a,                                      &                                      \\  \\
m<\frac{1}{\sqrt{2}}\,:                  & x^{(0)}.                                                      &
\end{array}
\end{equation}
 \subsection{ A calculation of a relative time delay}

As an example to show how to calculate relative time delays of explicit solutions of the lens equation   we consider
\begin{equation}\label{tid}
\tau(i y_{\pm}^{(2)})-\tau(0)=\frac{(y_{\pm}^{(2)})^2}{2}+m\,\Big(U(iy_{\pm}^{(2)})-U(0) \Big),
\end{equation}
where $i y_{\pm}^{(2)}$ (\ref{imas}) and $x^{(0)}=0$ are the bright images of $w=0$ in the two-cut case for $m>\frac{1}{\sqrt{2}}$ and $t\in(-m-\frac{1}{2m},-\sqrt{2})$. We observe  that
taking into account that
\begin{equation}\label{f0}
\partial_y U(iy)=-\int_{\Gamma}\rho(s)\frac{y}{y^2+s^2} d s,
\end{equation}
we have
\begin{equation}\label{f1}
U(iy)-U(0) =-\frac{i}{2}\int_0^y \Big(\omega(is)-\omega(-is) \Big)\, d s.
\end{equation}
Then, from (\ref{ccc}) and (\ref{f1}) we obtain
\begin{eqnarray}\label{f2}
\nonumber && \tau(i y_{\pm}^{(2)})-\tau(0)=\frac{(y_{\pm}^{(2)})^2}{2}-m\,\int_0^{y_{\pm}^{(2)}} \Big(s^3-t s-s\sqrt{(s^2-t)^2-2}\Big)\, ds\\
&&=\frac{1}{8m}\Big[1+4mt-4m^2\ln m+2m^2\left(1+t^2+t\sqrt{t^2-2}\right)\\
&&\quad-4m^2\ln\left(-t-\sqrt{t^2-2}\right)\Big].
\end{eqnarray}

\subsection{ The phase transition}

It is worth analyzing the behaviour of the images of $w=0$ at the phase transition  $t=-\sqrt{2}$ for the different values of the total mass $m$.

For $m>\frac{1}{\sqrt{2}}$
 it is easily found that
 \begin{equation}\label{p2}
\lim_{t\rightarrow-\sqrt{2}^+}x^{(1)}_{\pm}=\lim_{t\rightarrow-\sqrt{2}^-}x^{(2)}_{\pm},\quad
\lim_{t\rightarrow-\sqrt{2}^+}y^{(1)}_{\pm}=\lim_{t\rightarrow-\sqrt{2}^-} y^{(2)}_{\pm}.
\end{equation}
Thus the only effect on the set of images at the phase
transition is  the change from dim to bright  image of  $x^{(0)}=0$.

For $m=\frac{1}{\sqrt{2}}$ the image positions satisfy
$$\lim_{t\rightarrow-\sqrt{2}^+}x^{(1)}_{\pm}=\lim_{t\rightarrow-\sqrt{2}^+}x^{(d)}_{\pm}=\pm a,\quad
\lim_{t\rightarrow-\sqrt{2}^+}y^{(1)}_{\pm}=0,$$
so that the four bright images $x^{(1)}_{\pm}$, $iy^{(1)}_{\pm}$ which arise in the one-cut case  disappear at $t=-\sqrt{2}$.  Furthermore,  the dim images $x^{(d)}_{\pm}$ of the one-cut  case stay at $\pm a$ for $t<-\sqrt{2}$. As in the previous case, the dim image at $x^{(0)}=0$ in the one-cut case becomes bright in the two-cut phase.

 Finally, for $m\in(0,\frac{1}{\sqrt{2}})$ we have that according to (\ref{resonecut}) the images of $w=0$ for $t\in(-\sqrt{2},t_c)$ are $x^{(0)}=0$ (dim) and $iy^{(1)}_{\pm}$ (bright). Again
$$\lim_{t\rightarrow-\sqrt{2}^+} iy^{(1)}_{\pm}=0,$$
and these images disappear in the two-cut phase. Also as in the previous cases the dim image at $x^{(0)}=0$ becomes bright. These features of  the phase transition are shown in Fig. 11.
\begin{center}
\begin{figure}[ht]
\centering
\includegraphics[width=9cm]{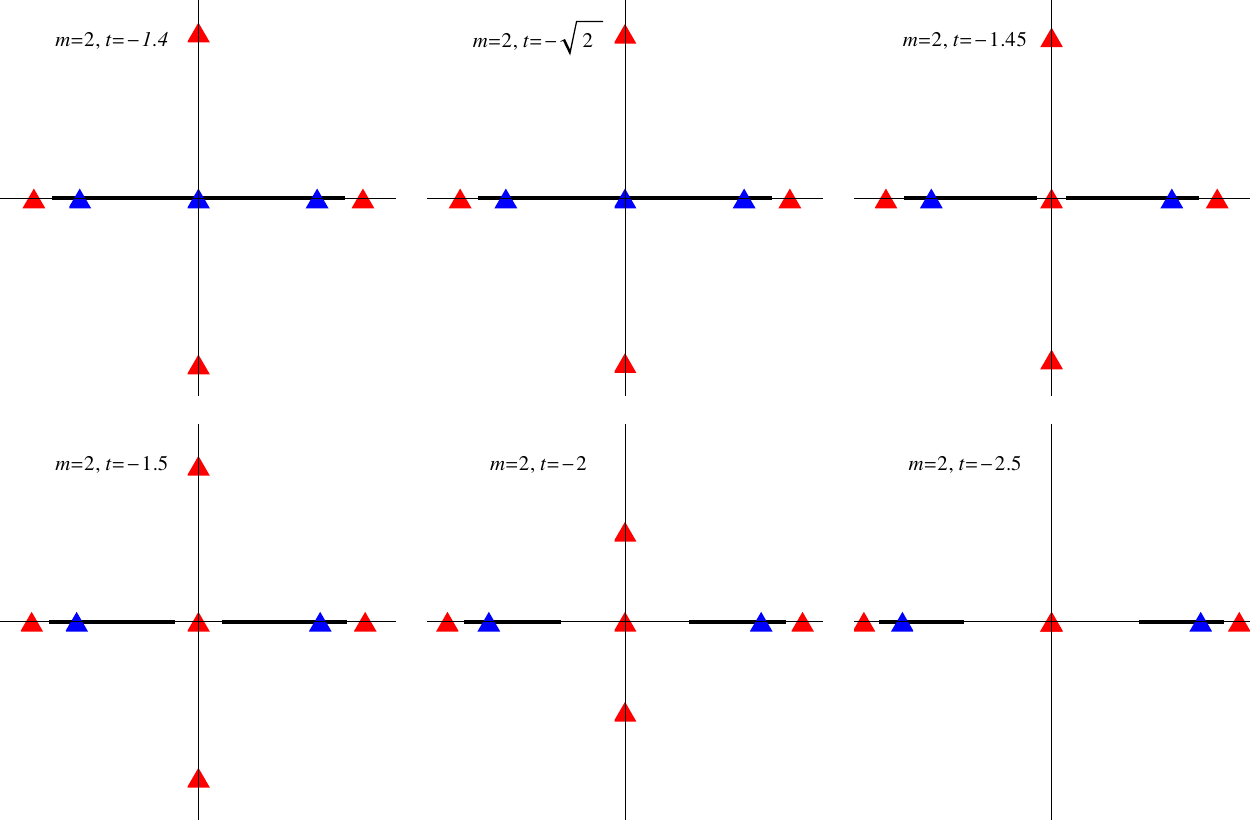}
\caption{Images for $w=0$ in the phase transition of the quartic model. The first figure corresponds to $t=-1.4$ just before the phase transition, we can observe an Einstein cross of images and a couple of dim images. Once the phase transition has occurred at $t=-\sqrt{2}$ (third to the sixth figures) the dim image at the origin has become bright. Moreover, the two bright images on the imaginary axis move towards the origin where they disappear at $t=-2.25$.
}
\end{figure}
\end{center}

  \section*{Concluding remarks \label{sec:con }}
In this paper we have  presented  several results  on the possible applications of the eigenvalue distributions of  random matrix models to gravitational lensing. We finish our discussion by raising several open problems.

From (\ref{ll}) it is clear that for a lensing model corresponding to a unitary ensemble with  potential $V(x)$ of degree $2p>2$  the number of dim images cannot exceed  $2p-1$ and that this bound is sharp. On the other hand, from (\ref{ll}) and taking (\ref{alg}) into account, Bezout theorem leads to the upper bound $4 p^2$ for the number of bright images. Although this bound is sharp for the Gaussian model it would  be interesting to improve it for the general case. 
 
 The present paper focuses on lensing models based on eigenvalue distributions of unitary ensembles of random matrices.  Nevertheless, lensing models with similar properties  can be generated from more general ensembles of random matrices \cite{AL13},\cite{AL15},\cite{AL17} in which the eigenvalues are constrained to lie on appropriate curves. It remains to know the interpretation of the associated lensing models.
 
 As it is well known \cite{DE99} the eigenvalue distributions of random matrices of large size  are closely connected to asymptotic distributions of zeros of associated families of orthogonal polynomials. Thus it is possible to formulate the analysis of this paper using zero distributions of orthogonal polynomials instead  of eigenvalue distributions of random Hermitian matrices. More generally, since the main  properties of these mass distributions derive from the fact that they minimize the  energy functional, we may  generalize our analysis to continuous critical measures in the sense of Mart\'{i}nez-Finkelstein and Rakhmanov   \cite{MA11}, which are saddle points for energy functionals.

  \section*{Appendix A: Eigenvalue distributions as mother bodies on elliptic domains \label{sec:app}}

 We next analyze  the two examples of mother bodies for measures supported on elliptic domains  which appear in this work.

 Let $D$ be the elliptic domain
 \begin{equation}\label{app2}
D=\{(x,y)\in \mathbb{R}^2,\, \frac{x^2}{\alpha^2}+\frac{y^2}{\beta^2}< 1,\,\alpha>\beta,\beta>0\}.
\end{equation}
In our next calculation  we  describe the  ellipse $\delta D$  in terms of the equation
 \begin{equation}\label{sch}
 \overline{z}=S(z),
 \end{equation}
where $S(z)$ is the Schwarz function of the ellipse \cite{SH92b}
\begin{equation}\label{sf0}
S(z)=\frac{\alpha^2+\beta^2}{a^2}z-i \frac{2\alpha \beta }{a^2}\sqrt{a^2-z^2},\quad (a^2:=\alpha^2-\beta^2).
\end{equation}
We also use the generalized  Cauchy formula
\begin{equation}\label{cgr}
\fl \int_{D} \frac{\overline{\partial}f(\zeta,\overline{\zeta})}{z-\zeta}\, d^2 \zeta=\pi \, n(\delta D, z)\,f(z,\bar{z})+\frac{1}{2 i}\int_{\delta D} \frac{f(\zeta,\overline{\zeta})}{z-\zeta} \, d \zeta,\quad z\in \mathbb{C}\setminus {\delta D},
\end{equation}
 for a smooth function  $f(z,\bar{z})$ on $\overline{D}$, where $n(\delta D, z)$ is the index of $z$ with respect to $\delta D$, $\overline{\partial}=\partial /\partial \overline{z}$ and $d^2 \zeta$ stands for the Lebesgue measure in the plane.

\subsection*{The Gaussian model}
Let  $\mu_D$ be the measure defined by  a uniform mass density $d \mu_D=d x \,d y$ on $D$. Then from (\ref{sch}) and (\ref{cgr}) it follows that its Cauchy transform  satisfies
\begin{equation}\label{swapb}
\omega(z)= \int_{D} \frac{d^2 \zeta}{z-\zeta}=\frac{1}{2 i}\int_{\delta D} \frac{S(\zeta)}{z-\zeta} d \zeta,
 \,\, z\in \mathbb{C}\setminus \overline{D}.
\end{equation}
If we deform $\delta D$ into the interval $[-a,a]$  and use the expression (\ref{sf0}) of $S(z)$ we obtain
\begin{equation}\label{swap3}
 \omega(z)=\frac{2 \alpha \beta}{a^2}\int_{-a}^a \frac{\sqrt{a^2-x^2}}{z-x}\, d x ,
 \quad z\in \mathbb{C}\setminus \overline{D}.
\end{equation}
The last integral is the Cauchy transform of the measure
\begin{equation}\label{dene}
d\mu_G(x)=\frac{2 \alpha \beta}{a^2}\sqrt{a^2-x^2}\,d x,\quad -a\leq x\leq a,
\end{equation}
which coincides with the mass distribution of the Gaussian model of total mass $m=\pi \alpha \beta$. Consequently,
$\mu_G$ is a mother body measure for $\mu_D$. In fact it can be proved \cite{SA05} that it is the unique mother body for $\mu$.

From  the generalized Cauchy formula,   the Cauchy transform of $\mu_D$ on $\mathbb{C}\setminus \delta D$  can be calculated \cite{FA09} in the form
\begin{equation}\label{swapa}
 \int_{D} \frac{d^2 \zeta}{z-\zeta}=
 \left\{\begin{array}{ll} 	\displaystyle  \pi \overline{z} +\frac{1}{2 i}\int_{\delta D}\frac{\overline{\zeta}}{z-\zeta} d \zeta
,\quad z\in D,
\\\\ \displaystyle  \frac{1}{2 i}\int_{\delta D} \frac{\overline{\zeta}}{z-\zeta} d \zeta,
 \quad z\in \mathbb{C}\setminus \overline{D},\ \end{array}\right.
\end{equation}
so that taking (\ref{sf0}) into account we obtain
\begin{equation}\label{swap2}
 \omega(z)=
 \left\{\begin{array}{ll} 	\displaystyle  \pi \overline{z} -\pi \frac{(\alpha-\beta)^2}{a^2}z
,\quad z\in D,
\\\\ \displaystyle  \frac{2\pi \alpha \beta}{a^2}\Big(z-\sqrt{z^2-a^2} \Big),
 \quad z\in \mathbb{C}\setminus \overline{D} .\end{array}\right.
\end{equation}

\subsection*{The quartic potential  model}
We now consider the  mass distribution of  the quartic potential model with mass $m$ in the one-cut case (\ref{qud1})
\begin{equation}\label{quda}
d \mu_Q(x)=\frac{m}{\pi}(x^2+c)\sqrt{a^2-x^2} \, d x,\quad -a\leq x\leq a,
\end{equation}
where the parameters $a$ and $c$ are given by (\ref{dens}).
Let $D$ be an elliptic domain (\ref{app}) with $\alpha^2-\beta^2=a^2$, and let us look for a measure $d \mu=\rho(z,\overline{z}) d^2 z$ on $D$ defining the same Cauchy transform as $d \mu_Q$ outside $D$. Assume a density of the form  $\rho=-i\overline{\partial} f$
 where $f=f(z,\overline{z})$ is a polynomial  such that its restriction to $\delta D$ is given by
\begin{equation}\label{q1}
f(z,\overline{z})=\frac{m}{\pi}(z^2+c)\sqrt{a^2-z^2}+g(z) ,\quad \forall z\in \delta D,
\end{equation}
where $g(z)$ is some polynomial in $z$. Then, from (\ref{cgr}) and deforming $\delta D$ into the interfocal interval $[-a,a]$ we have that the function $f$ must satisfy
\begin{equation}\label{q2}
-i\int_D  \frac{\overline{\partial}f(\zeta,\overline{\zeta})}{z-\zeta} d^2 \zeta=\int_{-a}^a \frac{d \mu_Q(x)}{z-x},\quad z\in \mathbb{C}\setminus \overline{D}.
\end{equation}

Taking into account that the expression (\ref{sf0}) of the Schwarz function for the ellipse can be written as $S(z)=A_1 z+i A_2 \sqrt{a^2-z^2}$ where
\begin{equation}\label{aes}
A_1=\frac{\alpha^2+\beta^2}{a^2},\quad A_2=-2\frac{\alpha \beta}{a^2},\quad a^2=\alpha^2-\beta^2,
\end{equation}
it is straightforward to see that a polynomial of the form
\begin{equation}\label{q3}
f(z,\overline{z})=\frac{m}{i \pi A_2}(c_1 \overline{z}^3 +c_2 z^2 \overline{z}+c_3 \overline{z}),
\end{equation}
verifies (\ref{q1}) if
\begin{equation}\label{syq}
(3 A_1^2+A_2^2) c_1+c_2=1,\quad c_3-a^2 A_2^2c_1=c.
\end{equation}
Moreover, (\ref{q3}) implies
\begin{equation}\label{dba}
-i\overline{\partial}f(z,\overline{z})=\frac{m}{\pi |A_2|}(3 c_1 \overline{z}^2+c_2 z^2+c_3),
\end{equation}
so that $-i\overline{\partial}f$ is a real-valued function if
\begin{equation}\label{syq2}
c_2=3 c_1.
\end{equation}
Thus from (\ref{syq}) and (\ref{syq2}) we deduce that a polynomial  of the form (\ref{q3}) satisfies (\ref{q1}) and  determines  a real-valued expression for $-i\overline{\partial}f$ provided
\begin{equation}\label{solu}
\fl c_1=\frac{1}{3 A_1^2+A_2^2+3},\quad c_2=\frac{3}{3 A_1^2+A_2^2+3},\quad c_3=c+\frac{a^2 A_2^2}{3 A_1^2+A_2^2+3}.
\end{equation}
In this case we have
\begin{equation}\label{pos2}
-i\overline{\partial}f (\zeta,\overline{\zeta})=\frac{m}{\pi |A_2|}\Big(2 c_2(x^2-y^2)+c_3\Big),
\end{equation}
so that for an elliptic region $D$ (\ref{app}) inside the open set
\begin{equation}\label{post}
y^2<x^2+\frac{c_3}{2 c_2},
\end{equation}
it follows that $-i\overline{\partial}f $ is positive on $D$. Then  $d\mu_D=-i\overline{\partial}f(z,\overline{z}) d^2 z$  determines a measure  supported on $D$. Moreover,  from (\ref{q2}) we have that the Cauchy transforms of $\mu_Q$ and $\mu_D$ coincide on $\mathbb{C}\setminus \overline{D}$,  and therefore it implies that  $\mu_Q$ is a mother body measure for $\mu_D$.

It is easy to see that a sufficient condition for $D$ to be contained inside the region (\ref{post}) is
\begin{equation}\label{suf}
\beta^2<\frac{a^2}{6}A_2^2=\frac{2}{3}\frac{\alpha^2\beta^2}{\alpha^2-\beta^2},
\end{equation}
or, equivalently,  $\alpha/\sqrt{3}<\beta<\alpha$.

\section*{Acknowledgments}

We wish to thank G. \'Alvarez for his help and many useful conversations. The financial support of the Spanish Ministerio de Econom\'{\i}a y
Competitividad under Project No. FIS2015-63966-P is also gratefully acknowledged.

\section*{References}
\providecommand{\newblock}{}

\end{document}